\documentclass[twocolumn]{aastex631}

\shortauthors{Belyakov et al.}
\begin{document}

\title{Limits on the Detection of Planet Nine in the Dark Energy Survey}

\correspondingauthor{Matthew Belyakov}
\email{mbelyako@caltech.edu}

\author[0000-0003-4778-6170]{Matthew Belyakov}
\affiliation{Division of Geological and Planetary Sciences, California Institute of Technology, Pasadena, CA 91125, USA}

\author[0000-0003-0743-9422]{Pedro H. Bernardinelli}
\affiliation{DiRAC Institute, Department of Astronomy, University of Washington, 3910 15th Ave NE, Seattle, WA, 98195, USA}

\author[0000-0002-8255-0545]{Michael E. Brown}
\affiliation{Division of Geological and Planetary Sciences, California Institute of Technology, Pasadena, CA 91125, USA}

\begin{abstract}

Studies of the clustering of the most distant Kuiper belt objects in the outer solar system have hinted at the possible existence of a planet beyond Neptune referred to as Planet Nine (P9). Recent efforts have constrained the parameter space of the orbital elements of P9, allowing for the creation of a synthetic catalog of hypothetical P9s. By examining the potential recovery of such a catalog within numerous sky surveys, it is possible to further constrain the parameter space for P9, providing direction for a more targeted search. We examine the ability of the full six years of the Dark Energy Survey (DES) to recover a synthetic Planet Nine population presented in \cite{brown2021orbit}. We find that out of 100,000 simulated objects, 11,709 cross the wide DES survey footprint of which 10,187 (87.0\%) are recovered. This rules out an additional 5\% of the parameter space after accounting for Planets Nine that would have been detected by both the Zwicky Transient Facility and DES.

\end{abstract}

\section{Introduction} \label{Sec:Intro}
The discovery of distant Kuiper belt objects (KBOs) on orbits that avoid Neptune's gravitational perturbations, starting with Sedna by \citet{sedna}, has spurred interest in scenarios of their origin and dynamical evolution. After the discovery of the Sednoid 2012 VP$_{113}$, \citet{2014Natur.507..471T} pointed out an apparent clustering of the argument of perihelion of these most distant planets around zero and speculated that this clustering could be
caused by a Kozai resonance with a distant planet, though no dynamical configuration that would cause such clustering could be found. Reexamining these objects, \citet{Batygin_2016} realized that the distant objects unaffected by Neptune are clustered in longitude of perihelion and pole position, rather than argument of perihelion. This cluster was found to be the natural outcome of a massive planet on an eccentric inclined orbit beyond Neptune. 

\begin{figure*}
\begin{centering}
    \includegraphics[width = 1 \textwidth]{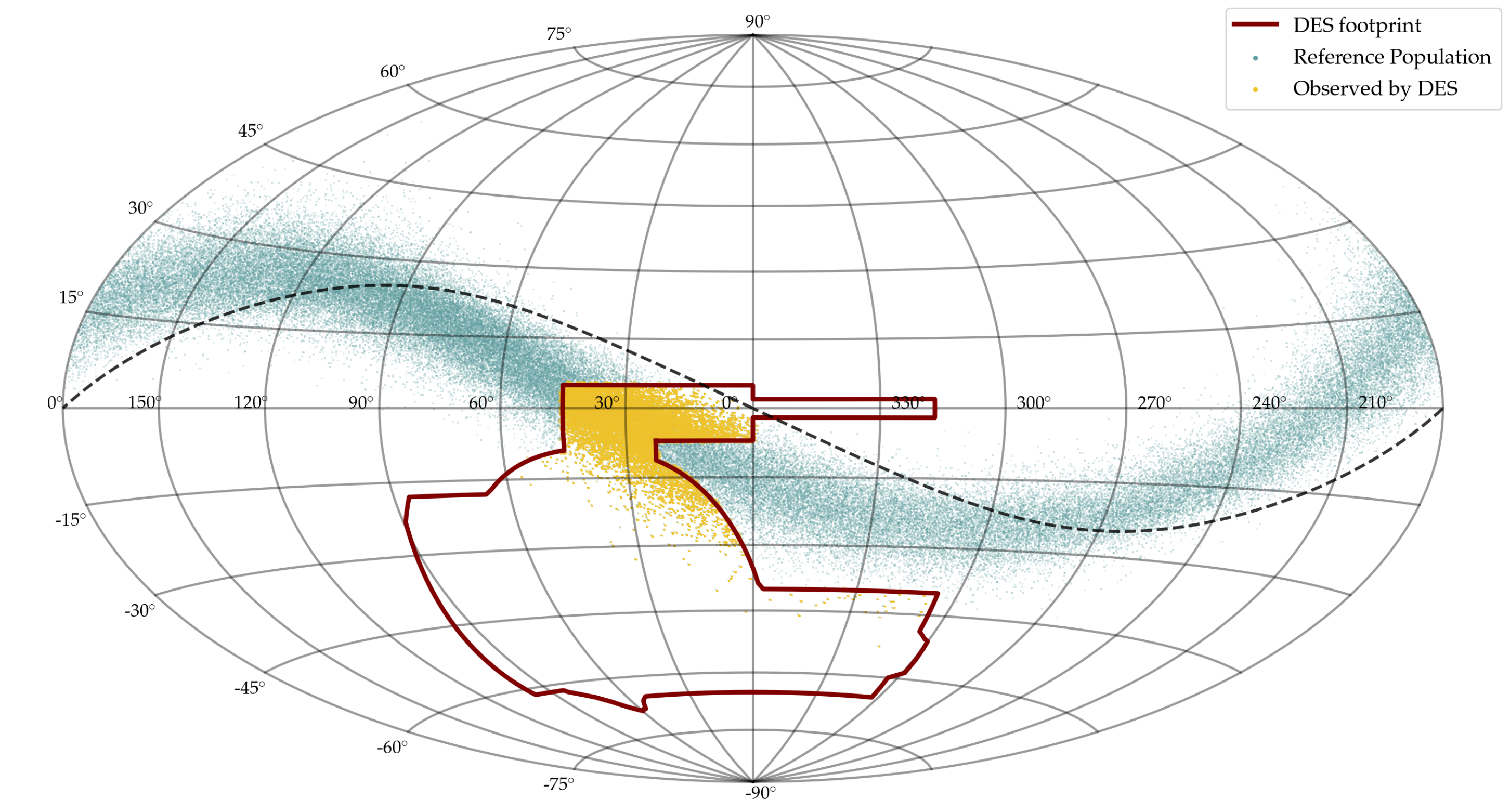}
    \caption{The synthetic Planet Nine population overlaid on top of the DES footprint. Blue dots represent the location of each member of the reference population on June 1st 2018. The yellow dots represent the individual detections of all members of the reference population which intersect the DES footprint. In red is the outline of the DES footprint, with the black dashed line tracing the ecliptic plane.}
    \label{fig:p9footprint}
    \end{centering}
\end{figure*}

Multiple lines of evidence now exist for the existence of a giant planet -- here called Planet Nine -- on a distant eccentric and inclined orbit, including the clustering in longitude of perihelion and in pole position of the highest semimajor axis KBOs, the existence of a substantial detached population of KBOs with perihelia removed far from the planetary region, and the abundance of Centaurs on orbits essentially perpendicular to the plane of the solar system \citep{2019PhR...805....1B}. While alternative explanations have been invoked to explain subsets of the phenomena \citep{2004AJ....128.2564M, 2009ApJ...697L..91G, shankman, Napier_2021, 2019AJ....157...59S, 2020PhRvL.125e1103S},  the existence of Planet Nine remains the simplest hypothesis unifying these disparate observations. Under the hypothesis that Planet Nine does exist, \cite{brown2021orbit} sought to provide direction for more targeted searches. The work combined information on observational biases with a likelihood model to constrain the orbital elements and mass of P9. Using the full orbital elements of 11 distant KBOs with $q>42$AU and associated observational biases, \cite{brown2021orbit} write a likelihood function for detecting each KBO with a certain ($i$, $\overline{\omega}$, $\Omega$) with known fixed semi-major axis and eccentricity for each KBO, given a Planet Nine with a certain mass, semi-major axis, and other orbital parameters. The product of these 11 likelihoods is then the full likelihood model for Planet Nine. From each of these likelihoods, a probability distribution for ($i$, $\overline{\omega}$, $\Omega$) assuming an isotropic distribution of KBOs was derived, and then multiplied by a biased probability distribution for each object. These probabilities were then converted into a final likelihood for Planet Nine, which was sampled using a Gaussian Process Emulator and Markov Chain Monte Carlo (MCMC) with two different prior choices, producing 49,100 uncorrelated samples. These samples capture the likely range of orbital parameters within which Planet Nine may be located, with a mass range of $m_9 = 6.2_{-1.3}^{+2.2}$ and semi-major axis range of $a_9 = 380_{-80}^{+140}$ \citep{brown2021search}. In order to aid the search for Planet Nine, this posterior sample for the parameter space was then sampled 100,000 times to create a synthetic catalog in \cite{brown2021search}. Effectively, each member of this catalog has a set of orbital elements that are randomly sampled from the parameter space derived in \cite{brown2021orbit}, along with a randomly assigned mean anomaly that places the simulated object in a specific place in its orbit. This catalog is therefore a reference population for P9, allowing it to be tested against existing surveys. Using radii derived from the sampled masses of these objects, the visual magnitudes of the synthetic P9s are found to range from 18 to 26, and thus are detectable by a multitude of sky surveys, such as Pan-STARRS, the Zwicky Transient Facility (ZTF), and Dark Energy Survey (DES). The ZTF search for Planet Nine in \cite{brown2021search} constrained 56\% of the parameter space for the orbital elements of P9 by detecting 56,173 members of the synthetic catalog injected into the ZTF survey.

The Dark Energy Survey is one of the deepest multi-epoch wide-field surveys to date and thus can provide useful constraints on the existence of Planet Nine. DES used the Dark Energy Camera \citep{decam} on the 4-meter Blanco Telescope in Cerro Tololo for 575 nights (2013–2019) to image a contiguous 5000 deg$^2$ patch of the Southern sky in five bands ($grizY$) over six years. Typical fields in the Wide Survey received around ten 90 second visits in each band in the 6 years, with typical limiting magnitudes of 23.8 in $r$ band. Although designed to probe the distribution of dark matter in the universe \citep{descollaboration2021}, the survey strategy has been proven to be capable of detecting hundreds of KBOs \citep{Abbott2016,Khain2020,Bernardinelli2019,bernardinelli2021search}. The final search for KBOs in DES \citep{bernardinelli2021search} yielded 814 objects, with magnitudes of 50\% completeness at around 24.71 mag in DES $g$-band and a well characterized detection efficiency. The northern-most latitudes that DES has imaged intersect with the simulated population generated in \cite{brown2021search}, allowing the survey to further constrain the P9 parameter space. With the public release of both the DES survey simulator \texttt{DESTNOSIM} and the accompanying Y6 data \citep{bernardinelli2021search}, a natural follow-up to \cite{brown2021search} is to test DES for its ability to recover the simulated Planet Nine catalog, as the DES magnitude limits suggest most members of the reference population can be recovered provided they intersect the DES footprint.

In this paper, we inject all 100,000 of the individual objects from the synthetic catalog of \cite{brown2021search} into the survey simulator for DES from \cite{bernardinelli2021search} for the full six years of the Dark Energy Survey to determine which parts of the parameter space for Planet Nine DES can exclude. We present the main results with assumptions for color and albedo identical to those of BB21. To supplement this analysis, we examine how different hypotheses on the potential surface type and composition of P9 can impact our ability to detect it. Finally, we demonstrate the effect of the combined results from DES and ZTF on the P9 parameter space.
\section{Simulated Planet Nine Population in DES} \label{Sec:main}
We use the P9 reference population created in \citep[][hereafter BB21]{brown2021search} to determine what further constraints DES can place on P9. In order to inject the simulated population into the survey, we must first make certain assumptions regarding the albedo, radius-mass relation, and colors of the reference population. For the main analysis, we adopt the same parameters from BB21. They used a uniform prior on albedo ranging from 0.2 to 0.75, with the lower end being half of Neptune's albedo, and the higher end derived from pure Rayleigh scattering based on \cite{2016}. The radius of any given synthetic object was derived from its mass, estimated by $r_9 = r_{\textrm{earth}}(m_9/3M_{\textrm{earth}})$. This relationship for the radius and mass comes from a best fit for Kepler data, which holds well in the mass range of Neptune \citep{2013ApJ...772...74W}. As DES observes in the $griz$ bands (note that \citealt{bernardinelli2021search} did not use the DES $Y-$band in their search), we must also make assumptions about the colors of the object. Using $g$-band as the reference band, we choose multiple models for the surface of Planet (listed in \autoref{tab:Table1}) to obtain a set of three colors. The fiducial model uses solar colors \citep{suncolor} derived from the assumption that P9's atmosphere will be a perfect Rayleigh scattering atmosphere \citep{2016}. This fiducial set of colors is also selected in order to match the colors assumed in the analysis from \cite{brown2021search}. The other four models chosen for the analysis are the model presented in \cite{2016} assuming a temperature of 40K and 10\% methane abundance, a Neptune-like P9, a ``super-Ganymede'', and a ``super-KBO''.
\begin{figure}
    \centering
    \includegraphics[width = 8.5cm]{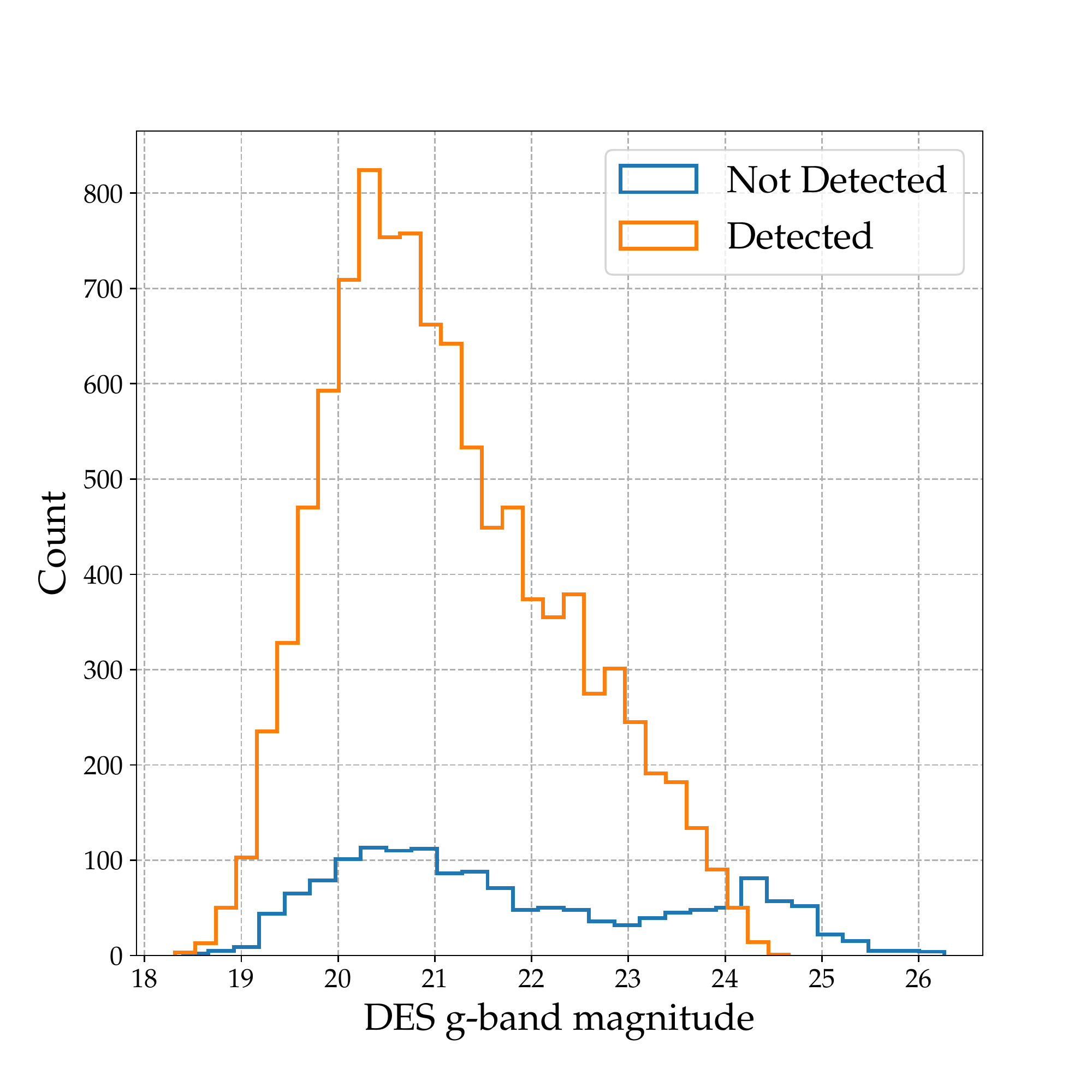}
    \caption{Distribution of the apparent $g$-band magnitudes of synthetic P9s in the DES footprint. Orange histogram shows the P9s that were successfully recovered, while the blue histogram shows the ones that were not recovered.}
    \label{fig:magdist}
\end{figure}

\subsection{Generating Synthetic Observations in DES} \label{Sec:obs}
We begin by determining which members of the synthetic population will intersect the DES footprint. Using the publicly available DES survey simulation software, \texttt{DESTNOSIM}\footnote{\url{https://github.com/bernardinelli/DESTNOSIM}} \citep{bernardinelli2021search}, we determine the detectability of our population of synthetic objects. The software accomplishes this task by finding which DES exposures intersect the object's orbit, followed by determining the likelihood of detecting the object individually in each exposure, and then checking against the recoverability of objects in the linking process. \cite{bernardinelli2021search} reported the search for KBOs to be $>99\%$ complete for objects at distances $29<d<2500$ AU (see also Figure 14 of \citealt{Bernardinelli2019}) that appear in at least 7 nights of data in at least 2 opposition seasons with at least a triplet of detections per season. We also use this set of criteria to determine which members of the synthetic population would have been detected.

For all synthetic objects, we use their orbital parameters to determine whether they intersect the footprint, and for those that do we obtain the RA/DEC coordinates and time of detection. In \autoref{fig:p9footprint}, we show the locations of all members of the synthetic catalog at the specified epoch of June 1st, 2018, in blue, with the detections simulated by DES in yellow. Note that only a small fraction of the survey footprint is intersected by the P9 search region.

In order to determine whether any given exposure would have detected an object with mass and color of any object in the reference population, we determine apparent magnitudes for each simulated P9. First, for each object in the reference population, we determine an apparent magnitude in DES $griz$ bands derived from its diameter $D$, assumed albedo $p$, colors relative to $g-$band, solar apparent magnitude $m_\odot$, and the object's distance from the sun and earth. The assumptions for colors and albedo are given in \autoref{tab:Table1}. In order to determine the apparent magnitude, we use the absolute magnitude of a body,
\begin{equation}
    H = m_\odot - 2.5\log_{10} \frac{pD^2}{9\cdot10^{16} \textrm{ km}^2}
\end{equation}
where the numerical factor is obtained from the definition \footnote{Starting from equation 5 and the following derivations in \cite{1916ApJ....43..173R}, convert to absolute magnitude -- a factor of (2 AU)$^2$ in kilometers gives the ${9\cdot10^{16} \textrm{ km}^2}$.} of geometric albedo. The absolute magnitude relates to the apparent magnitude as:

\begin{equation}
    m = H + 5\log_{10} (D_{ob} D_{sb}) - 2.5 \log_{10} \phi(\alpha) 
\end{equation}
where $D_{ob}$ is the observer-body distance and $D_{sb}$ is the sun-body distance. DES only observes at very small phase angles, so the phase function is $\approx 1$, thus combining the two equations, we obtain:

\begin{equation}
    m = m_\odot - 2.5\log_{10} \frac{pD^2}{9\cdot10^{16} \textrm{km}^2} + 5\log_{10} (D_{ob} D_{sb}) 
\end{equation}

In \autoref{fig:magdist}, we find that for solar colors and the albedos given in BB21, most members of the synthetic population are around 20-21st magnitude in DES $g-$band, with a long tailing end extending to 25th magnitude. 
\begin{figure}
    \includegraphics[width = 8cm]{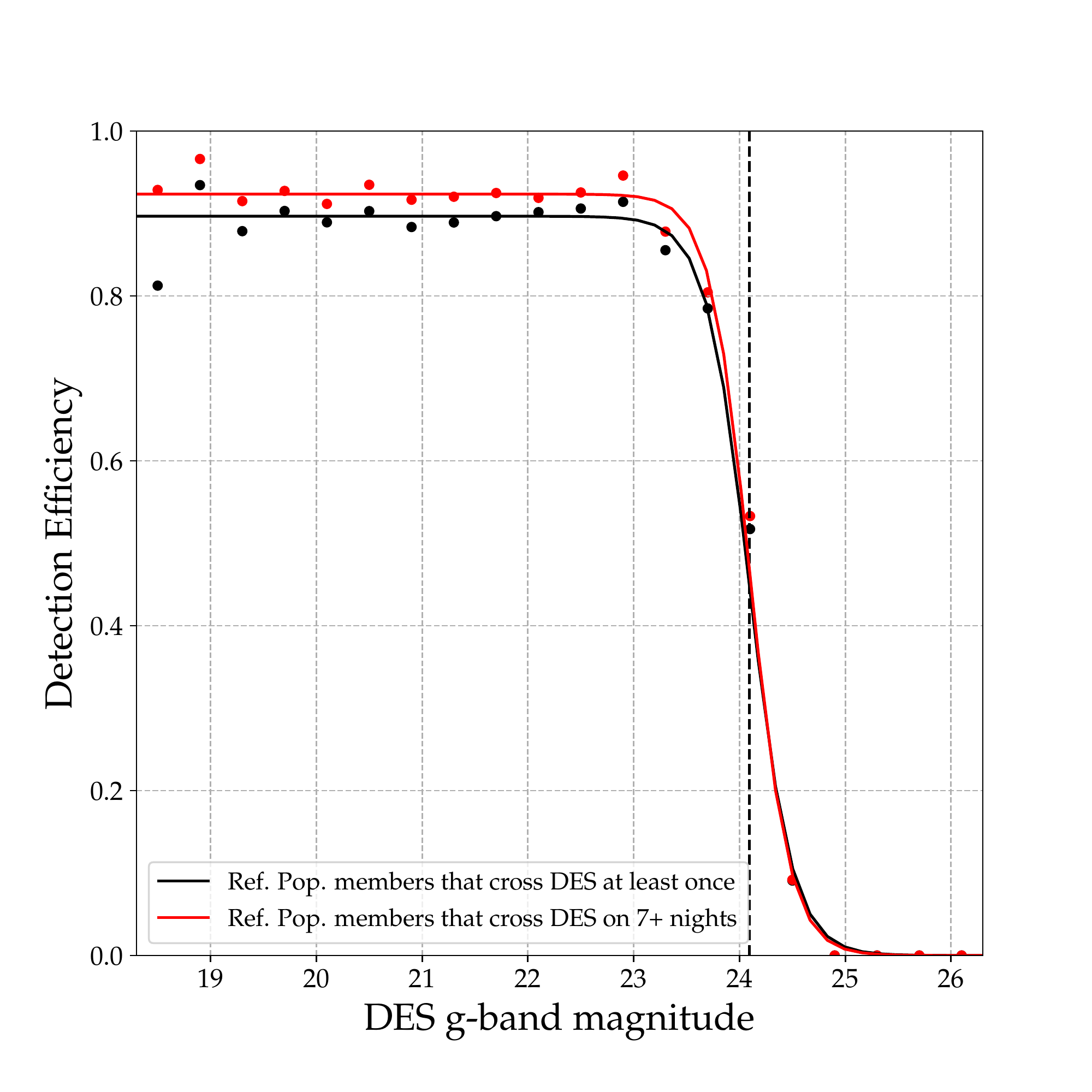}
    \caption{Completeness as a function of DES $g-$band magnitude for the whole population (black curve), as well as only the population with at least 7 possible detections, before any magnitude/transient efficiency cuts are imposed (red curve). Dashed line shows $m_{50}$ for both populations.}
    \label{fig:effi}
\end{figure}
\subsection{Recovery}
We now determine which synthetic observations would have actually been detected by checking the apparent magnitude of each detection against the limiting magnitude of each exposure. The completeness function for all exposures in DES was determined in \cite{bernardinelli2021search}, where the probability of a source being detected is modeled by a logit function:
\begin{equation}
    p(m) = \frac{c}{1+\exp(k(m-m_{50}))}
\end{equation} where $m$ is the detection's apparent magnitude, $m_{50}$ the magnitude where a source has a detection probability of 50\%, $k$ is a transition sharpness for the curve and $c \leq 1$ is a scaling factor that gives the detection efficiency for the brightest objects. This check is performed for every detection, substituting its apparent magnitude in for $m$, producing a probability of detection, unique for each detection and exposure. The $p(m)$ of each detection is checked against a uniform random number $u \in [0,1]$  -- if $p(m)<u$, the detection is discarded. Additionally, each detection is assigned another random number for the purposes of determining whether the detection would have been identified as a transient, which in DES was measured to be around $95.5\%$. This percentage includes transient identification using coadds ($98\%$ efficiency), treatment of sources with complex shapes misidentified as transients, and pixel masking, which when combined yield a total moving object detection efficiency. If a detection successfully passes both the transient check and the magnitude check, it is considered to be observed by the survey. 

In order to then determine whether the DES pipeline for finding TNOs would have detected members of the synthetic population, we apply the same recoverability criteria for DES Y6 Wide Field as \cite{bernardinelli2021search}:
\begin{enumerate}
    \item $\mathtt{NUNIQUE} \geq 7$. There must be more than 7 unique nights of detections.
    \item $\mathtt{ARCCUT} \geq 180$ days. The $\mathtt{ARC}$ is the time between the first and last detections in the orbit, and the $\mathtt{ARCCUT}$ is the shortest $\mathtt{ARC}$ that remains when any given detection is removed from the orbit. An $\mathtt{ARCCUT}$ of at least 180 implies there exist two detections at least 180 days after the rest.
    \item At least one triplet exists whose pairs are within 90 days of each other (as all of the objects here are farther than 50AU).
\end{enumerate}

Note that we do not fully inject these detections into DES -- we do not search for these objects using a full pipeline that would recover orbits. The detection efficiency determined in \cite{bernardinelli2021search} accounts not not only for incompleteness of the search at the faintest magnitudes, but also at the brightest ones. The suppression of detectability at 19-20 $g-$band mag is clearly visible in \autoref{fig:effi}, as detection of the reference population never reaches 100\%. This is due to a higher density of transients and stars in the northern part of the DES footprint, which is accounted for within the completeness function of each exposure, as $c<1$. Moreover, \texttt{DESTNOSIM} accounts for moving object retrieval by checking detections against the probability of transient detection, thus making it unnecessary to fully inject into the DES pipeline. Such injection would also be prohibitively expensive, as it would require the reprocessing of 80,000 sets of 62 CCD images, along with corresponding coadds. Additionally, the DES search was complete to a magnitude of $r \approx 23.8$ for distances $29<d<2500$ AU and 100\% efficiency for objects in the catalog above 200AU, which covers the entire range of semi-major axes relevant to the reference population \citep[see also][]{Bernardinelli2019}. For simplicity, we therefore assume that all P9 candidates that meet the three above criteria are recovered. Thus, as the efficiency of detection for distant KBOs has been determined for every exposure in DES by \cite{bernardinelli2021search}, it becomes simple to test for the completeness of P9 recovery in the survey.

\begin{deluxetable}{cccccc}
\tabletypesize{\footnotesize}
\tablewidth{0pt}
\tablecaption{Table showing percent detected of the reference population that crosses the DES footprint, given different albedos and colors selected based on various scenarios for the composition and surface of Planet Nine. \label{tab:Table1}}
\tablehead{\colhead{Model} & \colhead{Albedo} & \colhead{$g-r$} & \colhead{$g-i$} & \colhead{$g-z$} & \colhead{\% det}}
\startdata
Fiducial Model & BB21 & 0.44& 0.53& 0.55 & 87.0\%\\
Neptune-like & 0.5 & -0.3& -1.35& -2.15 & 80.3\% \\
40K, 0.1 CH$_4$ & 0.75 &0.6& -0.3& -0.2 & 88.1\% \\
Super-Ganymede & 0.43 &0.72& 0.88& 0.87 & 85.8\% \\
Super-KBO & 0.1 & 1.0& 1.25& 1.5 & 76.8\% \\
\hline
\enddata

\end{deluxetable}

\section{Results} \label{Sec:results}
We find that of the 100,000 objects in the P9 reference population, 11,709 cross the DES footprint as seen in \autoref{fig:p9footprint}, and of these, we are able to recover 87.0\%, or 10,187 objects. Many of the objects we fail to recover do not cross the footprint on six nights, even before transient efficiency or magnitude limits imposed by the detection probability per exposure are taken into account. This therefore suppresses the maximum efficiency at the lower magnitude range in \autoref{fig:effi}. In \autoref{fig:effi} we show the completeness of the P9 search in DES in DES $g-$band, finding an $m_{50}$ of around 24.1 apparent magnitude. This value is lower than the DES $g_{50}$ for a simulated TNO population determined in \cite{bernardinelli2021search}, likely due to differences in completeness for exposures in the ecliptic plane and unfavorable geometry in the northern latitudes of DES. In total, the extra depth of DES as compared to the ZTF search allows for the detection of 5032 objects not recovered by \cite{brown2021search} allowing us to rule out an additional 5\% of the reference population, and thus narrow possible parameter space for P9. It is worth noting that 98\% of the objects uniquely detected by DES have $g-$band magnitude greater than 20.75, which is the ZTF $m_{50}$.  
\begin{figure}
    \centering
    \includegraphics[width=8cm]{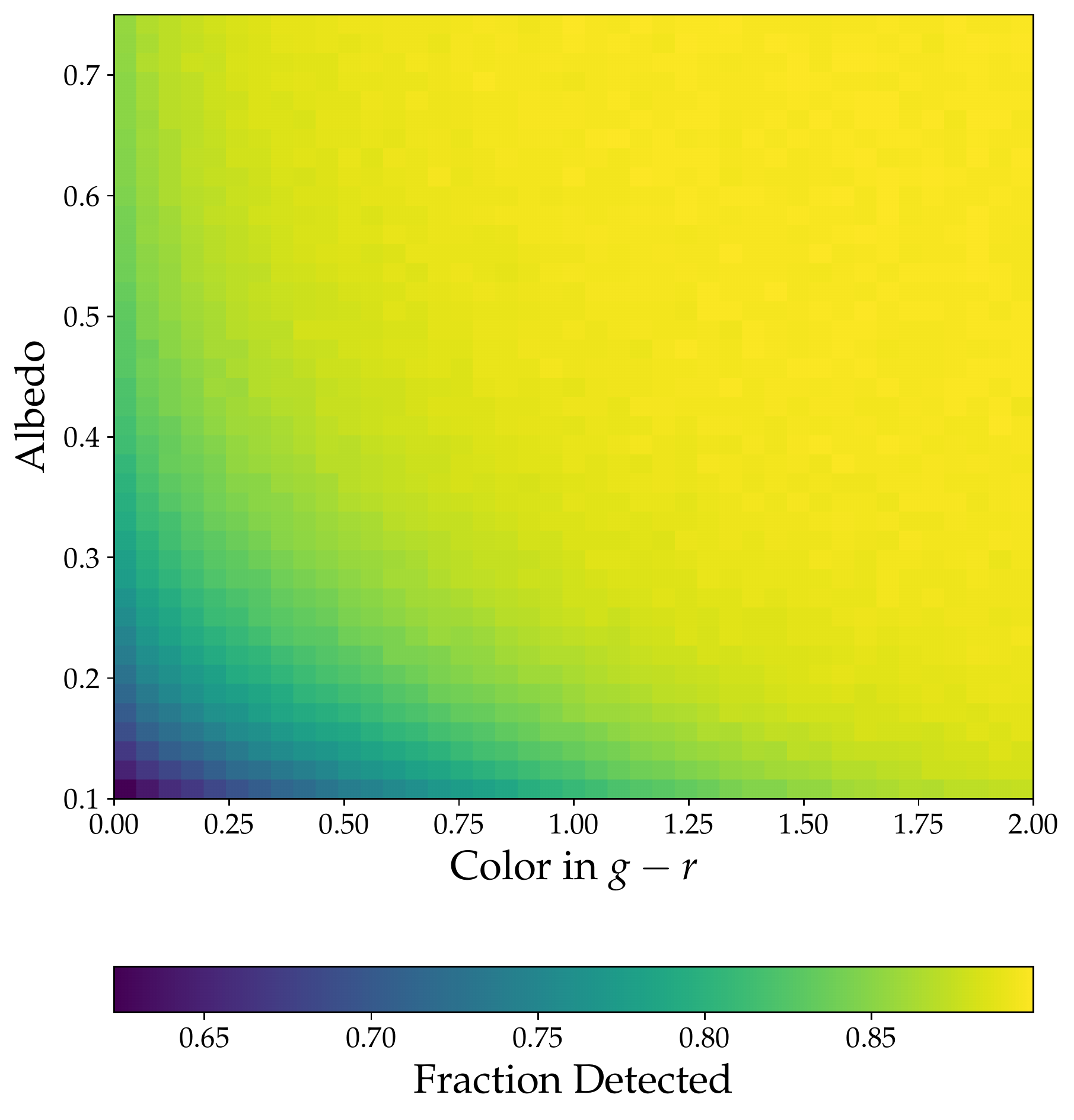}
    \caption{2D histogram in albedo vs. $g-r$ space, colored by the proportion of the reference population detected. Redder and higher albedo objects are more easily detected, while the bluer, low albedo range significantly depresses detection.}
    \label{fig:fracdet}
\end{figure}

While the standard synthetic population assumes solar colors and a range of albedos, we explore how different choices of color and albedo affect the detectability of the population. We repeat our analysis, now testing a spectrum of color and albedo instead of the BB21 values. We allow the $g-r$ color to vary from 0 to 2.0 and calculate the other two colors using the spectral slope defined by $g-r$. We also vary albedo from 0.1 to 0.75, thus creating a 2D space where for each combination of color and albedo we determine the proportion of the reference population detected. In \autoref{fig:fracdet} we plot the dependence of detectability on $g-r$ color and albedo. At the lower albedo, bluer range, we detect up to 25\% fewer members of the reference population, implying that a Neptune-like P9 would prove significantly harder to detect than a redder, higher albedo P9. 

We next explore how differing models for the nature of Planet Nine would affect the
detectability. 
First, we assume a Neptune-like scenario for P9, the results for which are given in the second row of \autoref{tab:Table1}. Substituting the exact values for albedo and color of Neptune while using the same fiducial model for the mass-radius relation, we recover 7\% fewer members of the reference population, dropping to 80\% recovery. The extremely blue colors of Neptune suppress recoverability despite the higher albedo, as confirmed by results in \autoref{fig:fracdet}. 

We next consider Neptune-like interior compositions, but more realistic atmospheres for a distant planet. The potential range of optical characteristics for P9 was analyzed in a series of models by \cite{2016}, in which assumptions for the temperature and methane abundance of P9 were used to determine its possible colors (see Table 2 of \citealt{2016}). Of these models we choose the one with a 40K temperature (the results in \cite{2016} do not significantly vary on temperature, and 40K is a middle-ground) and the maximum methane abundance. The resulting $g-r$ color is similar to the solar colors, however the other two colors are much bluer, though not nearly as blue as Neptune. We find that the effect of this model on the recovery of the synthetic population is minimal (third row of \autoref{tab:Table1}), and in fact increases the amount recovered by 1\% as compared to the fiducial model. The albedo from \cite{2016} is at the maximum of the range of albedos selected for the fiducial model, thus increasing the proportion of the population detected. 

\begin{figure}
    \centering
    \includegraphics[width=8cm]{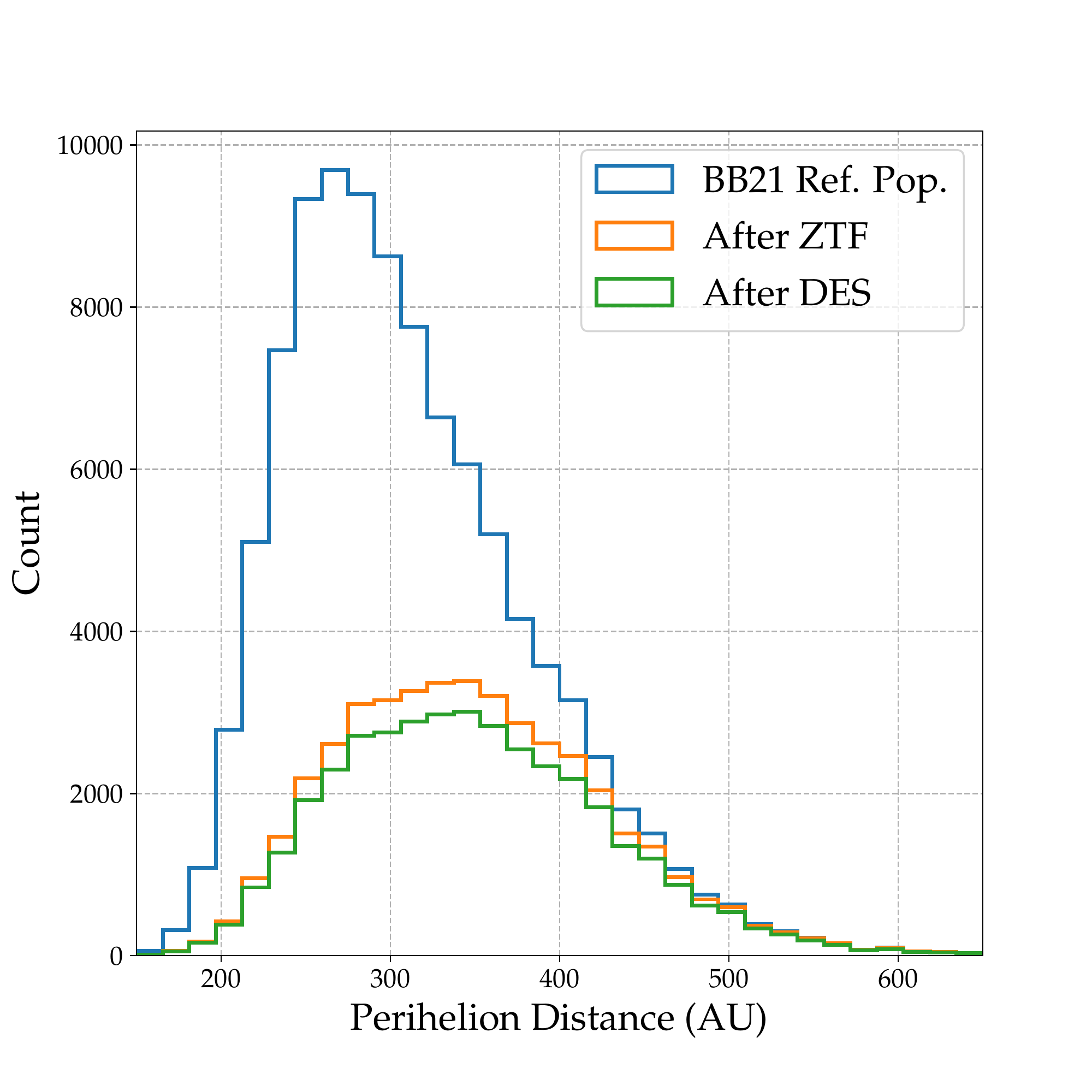}
    \caption{Histogram comparing the distribution of perihelion distance for the entire reference population in blue, the remaining population after the ZTF search in orange, and what remains after DES in green.}
    \label{fig:orbdist}
\end{figure}
For a very different end-member,  we examine a mixed ice-rock planet -- a ``super-Ganymede'' -- with composition similar to that of Ganymede at around equal parts rock and ice and with an albedo and color equal to those of Ganymede itself. We calculate the radius for each synthetic object by using an equation of state derived by \cite{seager} for a Ganymede-like planet. Effectively, this change narrows the extreme ends of the radius ranges for the reference population, with a 6 $M_\Earth$ planet yielding the same 2 $R_\Earth$ radius as previously
assumed, while a 10 $M_\Earth$ planet results in a radius of 2.27 $R_\Earth$ rather than 3.3 $R_\Earth$. We assign the ``super-Ganymede'' colors roughly estimated based on those determined in \citet{MILLIS1975408} and \citet{ Fadden1980VisibleSR}\footnote{We transform the $B-V$ color for Ganymede in \citet{MILLIS1975408} to SDSS $g-r$, and then use spectral data from \citet{Fadden1980VisibleSR} to extrapolate $g-i$ and $g-z$ colors.}. We find that a Ganymede-like P9 is not significantly different from the fiducial model in terms of ability to detect, as the albedo of 0.43 and very red color offset the smaller size of the planet. 

Finally, we consider a low albedo all-ice object that we term
a ``super-KBO.'' We assign this body generically red colors typical of classical KBOS, and give it an  albedo of 0.1. While the largest known KBOs have high albedos, we are
more interested in exploring a limiting case of one of the most difficult to detect
objects that we could envision.
We use the same equation of state from \cite{seager} as for the Ganymede-like P9, 
but with 100\% water ice (see Table 4 of \citealt{seager}). For this equation
of state that planet would be slightly larger than the super-Ganymede, such that a 6 $M_\Earth$ planet has a radius of 2.23 $R_\Earth$, while a 10 $M_\Earth$ planet goes up to 2.55 $R_\Earth$. At the higher mass end, these radii are smaller than those of the fiducial model, however are significantly larger at the lower end. Despite the increased radius, the lower albedo of the icy ``super-KBO'' model causes 10\% of the reference population to no longer be detectable. For both the KBO-like and Ganymede-like models, we find that the increase in radius at the lower mass ends offsets the recoverability loss from the decrease in radius at the higher mass end. We find a reduction in the fraction detected of only 10\%, from 87\% to just under 77\%.

\section{Conclusion} \label{Sec:conclusion}
We inject a simulated catalog of 100,000 objects generated in \cite{brown2021search} based on likely orbital parameters of Planet Nine into the Dark Energy Survey, recovering 10,187 of the synthetic objects (87\% of all objects that intersect the DES footprint), demonstrating that DES itself is capable of ruling out more than 10\% of the predicted parameter space of Planet Nine. Combining these results, DES and ZTF are able to rule out a significant portion of the reference population with perihelion distance under 300 as seen in \autoref{fig:orbdist}. We find that the median perihelion distance of the reference population after the two searches increases from 300 AU to 340 AU. However, DES itself does not significantly alter the distributions of orbital elements for the reference population as compared to ZTF, as it more evenly removes objects at all distances, eccentricities, and inclinations. The updated reference population, which now indicates which members of the population have been excluded by DES, can be found at \url{https://data.caltech.edu/records/11545} \citep{Brown_2022}. We find that modifying assumptions of the albedo and color of P9 can have a noticeable impact on its recoverability, such that a Neptune-like or KBO-like P9 would be somewhat more difficult to recover for surveys which lack the magnitude depth of DES. By contrast, assuming a Ganymede-like P9, or one similar to the models proposed in \cite{2016}, does not significantly alter our ability to recover members of the reference population. The methods used in this paper can be replicated for existing sky surveys, such as Pan-STARRS, Visible and Infrared Survey Telescope for Astronomy (VISTA), or SkyMapper to further constrain the parameter space for P9 and tighten the search. Additionally, the Rubin Observatory Legacy Survey of Space and Time (LSST) can be tested for its ability to recover the P9 reference population using simulated exposures in order to test how quickly the survey would be able to detect and rule out parts of the Planet Nine parameter space.

\bibliography{references}{}
\bibliographystyle{aasjournal}
\end{document}